\documentclass[preprint2]{aastex}

\shorttitle{Significance Tests and Background Estimation}
\shortauthors{Fleysher et al.}

\usepackage{subfigure}

\begin{document}

\title{Tests of Statistical Significance and Background
Estimation in Gamma Ray Air Shower Experiments.}

\author{R. Fleysher\email{roman.fleysher@physics.nyu.edu}
        L. Fleysher\email{lazar.fleysher@physics.nyu.edu}
        P. Nemethy\email{peter.nemethy@nyu.edu} 
        A. I. Mincer\email{allen.mincer@nyu.edu}}
\affil{Depatment of Physics, New York University, New York, NY 10003}

\and 

\author{T. J. Haines\email{haines@lanl.gov}}
\affil{Los Alamos National Laboratory, Los Alamos, NM 87545}

\begin{abstract}

In this paper we discuss several methods of significance calculation and
point out the limits of their applicability. We then introduce a self
consistent scheme for source detection and discuss some of its properties.
The method allows incorporating background anisotropies by vetoing
existing small scale regions on the sky and compensating for known large
scale anisotropies. By giving an example using Milagro gamma ray
observatory we demonstrate how the method can be employed to relax the
detector stability assumption. Two practical implementations of the method
are discussed. The method is universal and can be used with any large
field-of-view detector, where the object of investigation, steady or
transient, point or extended, traverses its field of view.
\end{abstract}


\section{Introduction}

The problem of evaluation of statistical significance of observations when
searching for gamma ray sources using air shower experiments remains one
of highest importance. The emission from a source would appear as an
excess number of events coming from the directions of the candidate over
the background level. The difficulty arises because the signal to
background ratio as registered by the detectors in this energy range is
often quite unfavorable, requiring careful examination of data.

In this paper, we expand on the usually adopted procedure of the
significance calculation described by \citet{lima}, in particular on the
conditions of its applicability. The prescription relies on the knowledge
of the expected background level, methods of estimation of which are
reviewed in \citet{cygnus_methods}. However, the standard significance
calculation method is not compatible with these methods of background
estimation. In this paper we introduce a self consistent scheme for a
source detection and discuss some of its properties. The method is
applicable to point and extended source searches as well as to searches
for transient phenomena. We show how practical problems specific to an
experiment can be incorporated into the method.

The methods described in this paper were developed for, and applied in two
gamma ray searches \citep{zorik_thesis,roma_thesis} using the Milagro
water Cherenkov air shower detector \citep{milagrito:nim}.

\section{Gamma ray astronomy using the air shower
technique.\label{section:air:shower:technique}}

A typical air shower detector registers particles from air showers that
survive to the ground level. The recorded information is used to provide
the direction of the incident primary particle and perhaps provide some
information on its energy and type. Among the particles entering the
Earth's atmosphere gamma rays present a very small fraction, often less
that $10^{-3}$. Most of the air showers are induced by charged cosmic rays
that form a background to the search for gamma initiated showers from a
source. Special techniques and algorithms have been developed to suppress
this background in order to increase the sensitivity to gamma primaries. 
These, however, are limited due to similarities of the cascades produced
by primaries of both types.  The application of these techniques helps but
does not solve the problem of gamma ray source detection in the presence
of a background. Therefore, one of the problems in gamma ray astronomy
using air shower technique is to be able to determine the level of
background \citep{cygnus_methods}.  This problem is rather difficult if
one tries to calculate it from the first principles, because it would
require exact knowledge of the details of the detector operation, its
sensitivity which may depend on voltages, temperature, properties of
atmosphere and direction reconstruction algorithms. The problem is solved
by measuring the background level using the same instrument. 

Thus, in a typical experiment, two measurements are performed --- one
corresponding to the observation of the candidate (so called {\em
on-source}) and the other is the measurement of the corresponding
background level (so called {\em off-source} observation). Then, a
decision is made as to the plausibility of the existence of the source.
Because the results of the on- and off-source observations represent
random numbers drawn from their respective parent distributions, the
question of the existence of the source is the question of whether the
numbers are drawn from the same or different parent distributions. It is
addressed by a hypothesis test.

\section{Li Ma statistic.}

Many statistical tests have been used to test the null hypothesis of the
absence of a source given two independent counts $N_{1}$ from the
on-source and $N_{2}$ from the off-source regions accumulated during time
periods $t_{1}$ and $t_{2}$ respectively with all other conditions being
equal. An improvement was proposed by \citet{lima} and is based on the
test statistic:

\begin{equation}
  U = \frac{N_{1} - \alpha N_{2}}{\sqrt{\alpha (N_{1} + N_{2}) }}
    \;\;\;\;  \alpha = t_{1}/t_{2} > 0
\label{equation:signif:lima}
\end{equation}

Because each event carries no information about another, each of the
observed counts can be regarded as being drawn from a Poisson distribution
with some value of the parameter (adjusted for the duration of
observation). The motivation provided by the authors is that the numerator
may be interpreted as the excess number of events from source over the
expected background and denominator is the maximum likelihood estimate of
the standard deviation of the numerator given that the null hypothesis is
true. 

It has been argued by the authors \citep{lima} that for the case of large
$N_{1,2}$ $(N_{1,2}>10)$, and if the null hypothesis is true, the
distribution of $U$ becomes Gaussian with zero mean and unit variance.
This statement is based on the well known fact that Poisson distribution
approaches that of Gaussian for large values of parameter $\mu$:

\[
  P_{\mu}(n) = \frac{\mu^{n}}{n!}e^{-\mu} 
             \approx
               \frac{1}{\sqrt{2\pi\mu}} e^{-{(n-\mu)^{2}}/{2\mu}}
\]

For a measured value $u$ of $U$, the calculation of the $p$-value (which
we denote by $\xi$) becomes simple: 

\[
 \xi = \frac{1}{\sqrt{2\pi}} \int_{u}^{+\infty} e^{-{x^{2}}/{2}} dx
\]

when looking for a source and

\[
 \xi = \frac{1}{\sqrt{2\pi}} \int_{-\infty}^{u} e^{-{x^{2}}/{2}} dx
\]

when looking for a sink. 

The null hypothesis is rejected with significance $\xi_{c}$ if $\xi <
\xi_{c}$. The significance $\xi_{c}$ is set in advance, before the test is
performed and its choice is based on the penalty for rejecting the null
hypothesis when it is true. (Scientific false discoveries should not
happen very often, and thus the significance is usually selected as
$\xi_{c}=10^{-3}$.) Because of the one-to-one correspondence between $\xi$
and $u$, the significance of a measurement can be quoted in the units of
$U$.

\subsection{Conditions of applicability of Li Ma statistic.}

\begin{figure}[tbp]
\centering
  \subfigure[Values of $N_{1}$ and $N_{2}$ are drawn from Poisson
             distribution with averages equal to 500 and 5000 
             respectively ($\alpha = 0.1$). According to equation 
             (\ref{equation:lima_bound}), $u_{b} \ll 5.3$.]{
    \label{fig:lima:a}
    \includegraphics[width=3.0in]{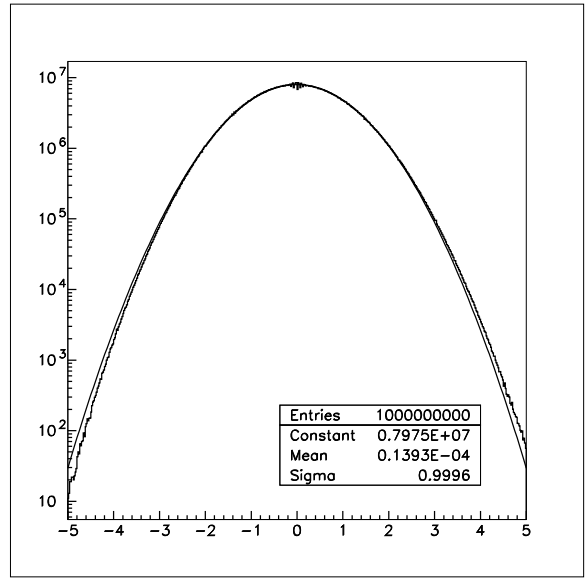}
  }
  \subfigure[Values of $N_{1}$ and $N_{2}$ are drawn from Poisson 
             distribution with averages equal to $15 \cdot 10^{5}$ and 
             $15 \cdot 10^{6}$ respectively ($\alpha = 0.1$). According
             to equation (\ref{equation:lima_bound}), $u_{b} \ll 20.3$.]{
    \label{fig:lima:b}
    \includegraphics[width=3.0in]{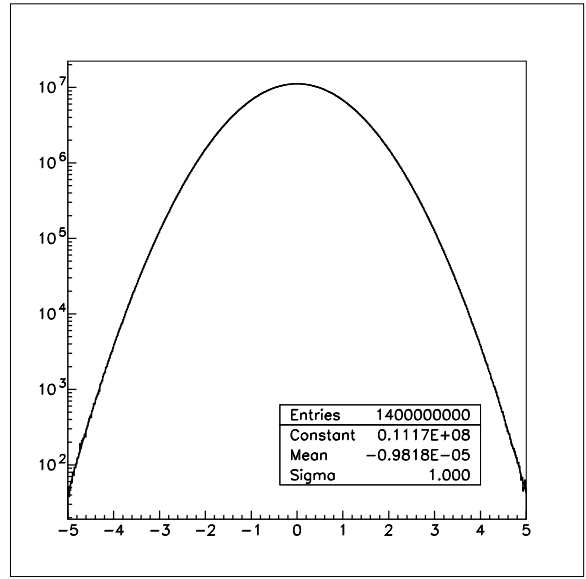}
  }

\caption{Distributions of statistic $U$ (equation
         \ref{equation:signif:lima}) obtained in two runs of Monte Carlo
         simulations. Parameters of the best fit to a Gaussian
         distribution are listed in the box. The two curves should agree
         in the $u_{b}$-neighborhood around zero. The number of entries in
         each run (about $10^{9}$) was chosen to provide reasonable
         accuracy in the region plotted.}
\label{fig:lima}
\end{figure}

\begin{figure}[tbp]
\centering
  \subfigure[Values of $N_{1}$ and $N_{2}$ are drawn from Poisson 
             distribution with averages equal to 500 and 5000
             respectively ($\alpha = 0.1$). According to equation 
             (\ref{equation:lima_bound}), $u'_{b} \ll 5.3$.]{
    \label{fig:notlima:a}
    \includegraphics[width=3.0in]{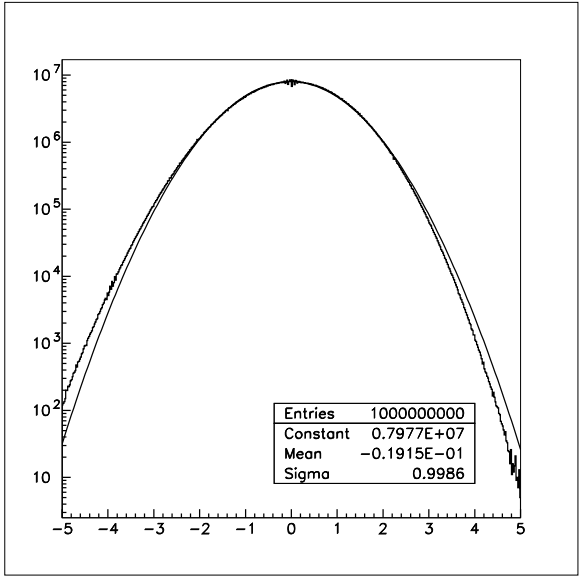}
  }
  \subfigure[Values of $N_{1}$ and $N_{2}$ are drawn from Poisson
             distribution with averages equal to $15 \cdot 10^{5}$ and    
             $15 \cdot 10^{6}$ respectively ($\alpha = 0.1$). According
             to equation (\ref{equation:lima_bound}), $u'_{b} \ll 20.3$.]{
    \label{fig:notlima:b}
    \includegraphics[width=3.0in]{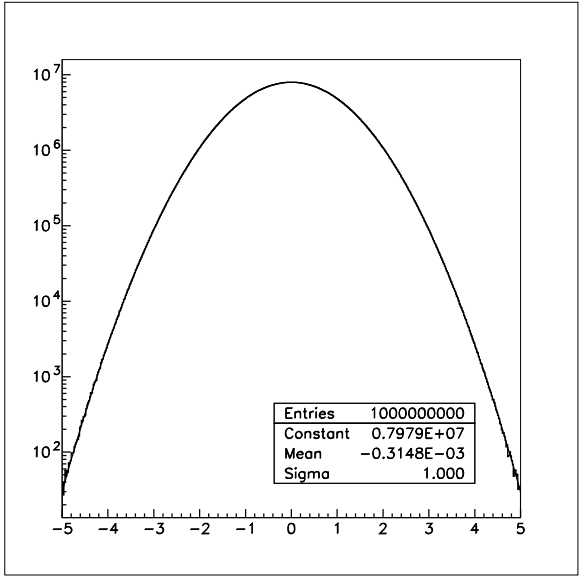}
  }

\caption{Distributions of statistic $U'$ (equation
         \ref{equation:signif:notlima}) obtained in two runs of Monte
         Carlo simulations similar to that presented in figure
         \ref{fig:lima}.}
\label{fig:notlima}
\end{figure}

When the null hypothesis is true, the distribution of statistic $U$
(equation \ref{equation:signif:lima}) approaches the normal distribution
in the limit of large numbers. Indeed, substituting the factorial in the
Poisson distribution using the Stirling formula $n! \approx \sqrt{2 \pi n}
\; n^{n} e^{-n}$, one obtains:

\[
  P_{\mu}(n) = \frac{\mu^{n}}{n!}e^{-\mu} 
             \approx \frac{1}{\sqrt{2 \pi n}} 
             \left(\frac{e \mu}{n}\right)^{n} e^{-\mu}
\]

or

\[
  \ln \left( P_{\mu}(n)  \sqrt{2 \pi n} \right) \approx
       n \left( 1 - \ln \frac{n}{\mu} \right) - \mu
\]

Expanding the right hand side into Taylor series in $n$ in the vicinity of
$\mu$, denoting $\delta = (n -\mu)$ and keeping first two non-zero terms,
we obtain: 

\[
  P_{\mu}(n) = \frac{\mu^{n}}{n!}e^{-\mu}
             \approx \frac{1}{\sqrt{2 \pi \mu (1 + \delta / \mu) }}
            e^{-{\delta^{2}}/{2\mu}} 
            \cdot e^{{\delta^{3}}/{6\mu^{2}}}
\]

Thus, it is seen that the Poisson distribution approaches that of Gauss in
a narrow region around its mean: $\left| {\delta^{3}}/{6\mu^{2}} \right| =
\left| {(n-\mu)^{3}}/{6\mu^{2}} \right| << 1$ with $\mu \ge 6$.
Substituting $n$ with $N_{1,2}$ and $\mu$ with corresponding estimates of
$\mu = (N_{1}+N_{2}) t_{1,2}/(t_{1}+t_{2})$ we obtain the region around
zero $|u| < |u_{b}|$ where the distribution of statistic $U$ is
approximately normal. That is, for

\begin{eqnarray}
 \lefteqn{ |u| < |u_{b}| << }  \nonumber \\
        & \min & \left( 
                 \sqrt[6]{36\alpha     (1+\alpha)^{2}(N_{1}+N_{2})}, \; 
                 \right. 
                 \label{equation:lima_bound} \\
             & & \left.
                 \sqrt[6]{36\alpha^{-3}(1+\alpha)^{2}(N_{1}+N_{2})}
                 \right) \nonumber
\end{eqnarray}

the error on the $p$-value $\xi$ due to this approximation does not exceed
$ {1}/{\sqrt{2\pi}} \; \int_{u_{b}}^{+\infty} e^{-{x^{2}}/{2}} dx$.
Figure \ref{fig:lima} shows the results of a Monte Carlo simulations for
distribution of the statistic $U$. It can be seen that the distribution is
approximately normal in the vicinity of zero.

By the same arguments it may be shown that within essentially the same
region around zero, another statistic $U'$

\begin{equation}
  U' = \frac{N_{1} - \alpha N_{2}}{\sqrt{N_{1} +\alpha^{2}  N_{2}}}
\label{equation:signif:notlima}
\end{equation}

is also distributed normally. The motivation for statistic $U'$ is similar
to that of statistic $U$ (equation \ref{equation:signif:lima}) that the
numerator may be interpreted as the excess number of events from source
over the expected background but the denominator is the maximum likelihood
estimate of the standard deviation of the numerator given that the
alternative hypothesis is true. (The alternative hypothesis in this case
is that both observations $N_{1}$ and $N_{2}$ are from Poisson
distributions with unrelated means.) Although this motivation appears to
be incorrect and the statistic was abandoned by \citet{lima}, the critical
range of $U'$ may be defined as $u'>u'_{0}$ for testing the null
hypothesis against the presence of a source and $u'<u'_{0}$ for testing
against the presence of a sink. Because under the conditions of
applicability of Li Ma statistic (equation \ref{equation:lima_bound}) 
statistic $U'$ is distributed normally, the $p$-value calculation is
identical to that of for statistic $U$. The figure \ref{fig:notlima}
presents the results of Monte Carlo simulations of distribution of
statistic $U'$. 

In general, a hypothesis test may be based on any statistic if its
distribution under the null hypothesis in known. 

It is interesting to note that equation (\ref{equation:lima_bound}) can be
used to aid in the design of an experiment. Indeed, if the relative on- to
off-source region exposure can be estimated before the experiment is
performed, then equation (\ref{equation:lima_bound}) allows estimating the
observation time needed to collect enough events to reach the accuracy of
statistics $U$ or $U'$ compatible with the desired significance
$\xi_{c}$. For example, if $\alpha \approx 0.1$ and the significance in
units of $U$ is set at 3.0, then the experiment (duration of observation) 
has to be designed in such a way as to allow accumulation of at least
$10^{5}$ events from the on-source region. If the significance is set at
5.0, then number of on-source events should be at least $10^{6}$.

\section{Background estimation.}

In order to be able to implement any of the above hypothesis tests, one
must assure that the two measurements $N_{1}$ and $N_{2}$ are independent
and that the ratio of observation times $\alpha$ is available while other
conditions are equal. Indeed, examining a typical scenario of a gamma ray
experiment it is seen that on- and off-source observations can be
performed at the same time utilizing the wide field of view of the
detector, or they can be performed at different times making measurement
in the same local directions of the field of view. (Due to the Earth's
rotation, the off-source bin may present itself in the directions of local
coordinates previously pointed at the source bin.) Both of these
stipulations could contradict to the conditions of ``being equal'': if
observations are done at the same time, then non-uniformity in the
acceptance of the array to air showers due to detector geometry must be
compensated for; if observations are done at different times, then any
time variation in detector operation must be addressed. Under these
varying conditions, the meaning of the parameter $\alpha$ must be changed
to the effective ratio of exposures of the bins. The mechanism of such an
equalization and $\alpha$ determination is called {\em background
estimation}. The name is due to interpretation of the second term of the
numerator of equation (\ref{equation:signif:lima}) as the expected number
of background events in the source region: $N_{b} \equiv \alpha N_{2}$.
Correspondingly, the number of events $N_{1}$ obtained from the direct
source observation will be denoted $N_{s} \equiv N_{1}$. Below we consider
two methods of background estimation: direct integration and time
swapping.

\subsection{Isotropy and stability assumptions.}

A widely accepted method of background estimation \citep{cygnus_methods}
recognizes that usually no major changes in the detector configuration are
made on short time scales and takes advantage of the rotation of the
detector with the Earth which sweeps the sky across the detector's field
of view. It also recognizes that most air showers detected are produced by
charged cosmic rays. Because of their charge and because of the presence
of random magnetic fields in the interstellar medium, the cosmic ray
particles lose all memory of their initial directions and sites of
production, and can be regarded as forming isotropic radiation. Detector
configuration stability implies that the acceptance of the detector is
time independent although variations in the overall rate of detected
events are allowed. (An example of such rate variations could be an event
rate decrease caused by a temporary data acquisition system overload.) 
Therefore, the average number of detected events as a function of local
coordinates $x$ and time $t$ on the short time scale can be written in the
form: 

\begin{equation}
 d N(x,t) = G(x) \cdot R(t) \; dx dt
\label{equation:sloshing:stability_assumption}   
\end{equation}

Here $R(t)$ is overall event rate, $G(x)$ --- acceptance of the array such
that $\int_{field \; of \; view} G(x) \; dx = 1$. The local coordinates
$x$ could be either hour angle and declination or zenith and azimuth. The
average number of background events expected in the source bin, is then
given by

\begin{equation}
 N_{b} = \int \int (1-\phi(x,t)) \; G(x) \; R(t) \; dx dt
\label{equation:sloshing:background}
\end{equation}

where $\phi(x,t)$ is equal to zero if $x$ and $t$ are such that they
translate into inside of the source bin, and is one otherwise. The
isotropy and stability assumptions (equation
\ref{equation:sloshing:stability_assumption}) become part of the null
hypothesis being tested.

\subsection{Direct integration method.}

The direct integration method of source detection is based on isotropy and
stability assumptions (equation
\ref{equation:sloshing:stability_assumption}) and is the method where the
integration of equation (\ref{equation:sloshing:background}) is performed
numerically by discretizing both $G(x)$ and $R(t)$ on a fine grid and
replacing integrals by sums. The significance test is based on either
statistic (\ref{equation:signif:lima}) or (\ref{equation:signif:notlima}).
The acceptance and the event rate are estimated by histogramming local
coordinates $x$ and event times $t$ of the events collected during
integration time period from the entire sky. The fluctuations in $N_{b}$
are dominated by the ones in $G(x)$ because the event rate $R(t)$ is
collected from the entire sky and may be deemed as known to high
precision. In this scheme, the source region defined by $\phi(x,t)$ also
gets discretized, therefore, source count $N_{s}$ must be obtained using
the same discretized definition of the source region. Extending the time
integration window is equivalent to increasing exposure to the off-source
bin, which leads to decreasing value of $\alpha$ and improved sensitivity.
The assumption (\ref{equation:sloshing:stability_assumption}), however,
must hold during the entire integration period placing a constraint on the
maximum size of the off-source bin. The time integration window is limited
by 24 hours of sidereal day.

\subsubsection{Eliminating on-source events from background estimation.
                        \label{section:background:signal_events}}

The realization of the direct integration method just described includes
on-source events $N_{s}$ in the calculation of expected background $N_{b}$
(via $G(x)$ and $R(t)$). This, however, is inconsistent with their
independence required by Li Ma statistic (equation
\ref{equation:signif:lima}) and was already recognized in
\citep{cygnus_methods}. An extreme example would be the case of sighting
of the North Celestial Pole. There, the source does not present any
apparent motion in local coordinates because it lies on the axis of
rotation of the Earth. The off-source bin does not exist, the off-source
count $N_{2}$ and the ratio of exposures $\alpha$ are not defined and the
measurement can not be performed using isotropy and detector stability
assumptions (equation \ref{equation:sloshing:stability_assumption}). In
the framework of the direct integration method, however, the background
$N_{b}$ is guaranteed to be estimated exactly equal to $N_{s}$ and
therefore $U \equiv 0$. This is clearly unsatisfactory. 

In order to be able to use either of the statistics
(\ref{equation:signif:lima}) or (\ref{equation:signif:notlima}) the events
from the source bin should be excluded from the background estimation.
However, simply removing all of these events from the procedure will
destroy its foundation that the lists of local coordinates and times
represent samples from $G(x)$ and $R(t)$ respectively. A solution to this
problem follows.

Denote by $\psi(x,t)$ a function similar to $\phi(x,t)$ which defines the
region of the sky events from which are to be excluded from the background
estimation. The excluded region should contain the candidate source bin,
but is not limited to it. Also denote by $N_{out}(x,t)$ the number of
detected events originating from outside of the excluded region, and
$R_{out}(t)$ their total event rate, then it is readily seen that

\[
  d N_{out}(x,t) = \psi(x,t) G(x) R(t) \; dx d t
\]

Integrating this equation with respect to $t$ and $x$ a system of
equations on unknown $G(x)$ and $R(t)$ is obtained ($N_{out}(x)$ and
$R_{out}(t)$ are available experimentally):

\begin{equation}
  \left\{
  \begin{array}{lcl}
   N_{out}(x) & = & G(x) \int \psi(x,t') \cdot R(t') \; dt'  \\
   R_{out}(t) & = & R(t) \int \psi(x',t) \cdot G(x') \; dx'
  \end{array}
  \right|
\label{equation:background:equations}
\end{equation}

\begin{figure}[t]
\centering
\includegraphics[width=3.0in]{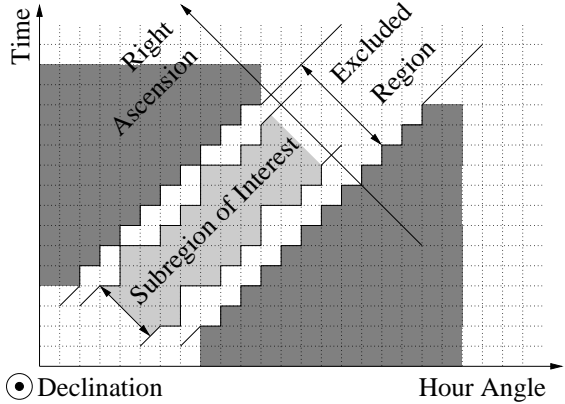} 
\caption{To the formulation of background equations
         (\ref{equation:background:equations}).}
\label{fig:background:solution}
\end{figure}

The numerical solution of these integral equations provides $R(t)$ and
$G(x)$ based on data $N_{out}(x)$ and $R_{out}(t)$ from the outside of the
excluded region to be used in equation
(\ref{equation:sloshing:background}). The situation is illustrated on
figure \ref{fig:background:solution}. The heavily shaded area is the
outside of the excluded region bounded by $\psi(x,t)$ in its discrete
form, events from which may be used for the off-source observation. The
region of interest, the on-source region, is defined by some other
conditions $\phi(x,t)$ which are irrelevant for the background equations
(\ref{equation:background:equations}) as long as it is contained in the
excluded region.


It can be noted that both $G(x)$ and $R(t)$ enter into equations
(\ref{equation:background:equations})  and
(\ref{equation:sloshing:background}) only as a product $G(x) \cdot R(t)$,
therefore, normalization of either of them does not make any difference as
long as the product is preserved. Also, if there are points $\{x_{0}\}$ in
the local coordinates which are always inside the excluded region, which
may happen if the detector was operational during a short time period
and/or the excluded region was large, (that is $\psi(\{x_{0}\},t) = 0, \; 
\forall t$) then $N_{out}(\{x_{0}\}) \equiv 0$ and the first equation of
(\ref{equation:background:equations}) becomes: 

\[
    0 = G(\{x_{0}\}) \cdot 0
\]

leading to $G(\{x_{0}\})$ and integral in equation
(\ref{equation:sloshing:background}) being undefined. On-source events
with local coordinates from these regions must be discarded as having no
corresponding background estimate. It is thus seen that the method entails
that the second, off-source region is defined by the regions of the {\em
local} sky which have the opportunity to present themselves into the
directions of the source region due to the Earth's rotation during the
time period of integration. Different parts of the source region have
different corresponding off-source regions. This leads to the ratio of
exposures of on- and off-source regions $\alpha$ that is dependent on the
local coordinates $x$: 

\[
  \alpha(x) = \frac{N_{b}(x)}{N_{out}(x)} 
            = \frac{\int (1-\phi(x,t)) R(t) dt}{\int \psi(x,t) R(t) dt}
\]

The off-source region corresponding to a given on-source region
$\phi(x,t)$ is not a celestial bin, it is a set of local directions with
$\alpha(x) > 0$. Because the measurements made from different local
directions $x$ are independent, all measurements can be combined to obtain
the compound statistic $U$:

\[
  U=\frac{\sum_{x}N_{s}(x) -
    \sum_{x}N_{b}(x)}{\sqrt{\sum_{x} \alpha(x)N_{s}(x)+\sum_{x}N_{b}(x)}}
\]

or

\[
  U=\frac{N_{s} - N_{b}}{\sqrt{\sum_{x} \alpha(x)N_{s}(x)+N_{b}}}
\]

The described method is the integration scheme which is based on the
direct integration method and which properly estimates the ratio of
exposures $\alpha(x)$ and accounts for the source events.

\begin{figure}[tbp]
\centering
  \subfigure[Excess number of events $(N_{s}-N_{b})$ as a function of
             Galactic Latitude. Source region is not excluded.]{
    \label{fig:galactic_latitide:mc:before}
    \includegraphics[width=3.0in]{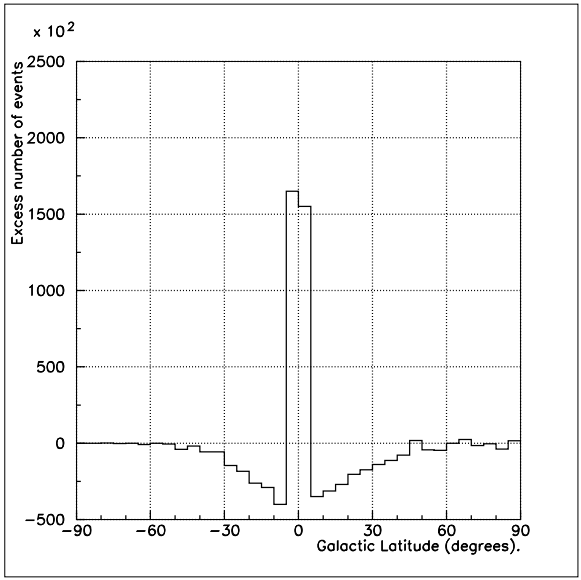}
  }
  \subfigure[Excess number of events $(N_{s}-N_{b})$ as a function of
             Galactic Latitude. The region of $\pm 7^{\circ}$ around    
             Galactic equator is excluded from background estimation.]{
    \label{fig:galactic_latitide:mc:after}
    \includegraphics[width=3.0in]{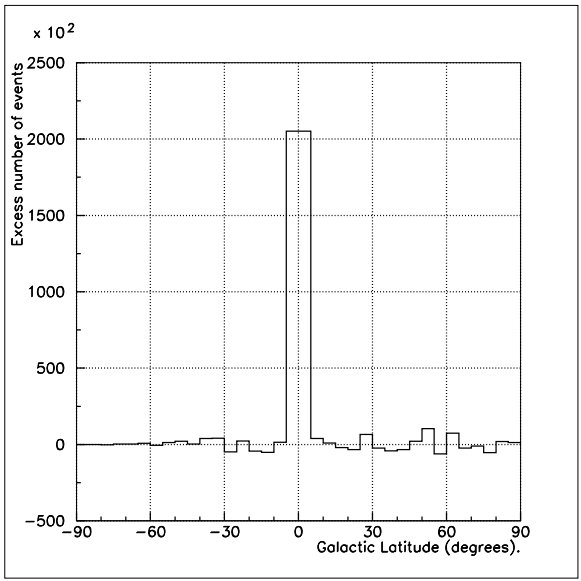}
  }

\caption{Plots showing the results of Monte Carlo simulations with uniform
         Galactic signal flux being 0.0088 that of background in the
         region of $\pm 5^{\circ}$ around the Galactic equator. The
         expected Galactic bin content is about 205000.}
\label{fig:galactic_latitide:mc}
\end{figure}

The importance of the source region exclusion is illustrated on figure
\ref{fig:galactic_latitide:mc} where results of the computer simulations
for a Galactic plane observation is presented. Detection of an extended
source such as Galactic plane presents a difficult example because the
ratio of on- and off-source exposures $\alpha(x)$ varies dramatically over
the area of the source. The figure shows the excess number of events
$(N_{s}-N_{b})$ extracted from a simulated galactic signal as a function
of Galactic latitude. The excess is recovered correctly by the modified
method (equations \ref{equation:sloshing:background} and
\ref{equation:background:equations}) proposed here (figure
\ref{fig:galactic_latitide:mc:after}) compared to the standard direct
integration method (equation \ref{equation:sloshing:background}) (figure
\ref{fig:galactic_latitide:mc:before}). Use of the standard method would
lead to a 25\% loss in both the excess number of events and in value of
statistic $U$ which are recovered by the modification.

\subsection{Time swapping method.}

In the time swapping method of source detection the integration of
equation (\ref{equation:sloshing:background}) is performed by means of
Monte Carlo, which leads to: 

\begin{equation}
  N_{b} = \frac{N_{0}}{N} \sum_{i=1}^{N} (1-\phi(x_{i},t_{i}))
        = \frac{1}{\beta} \sum_{i=1}^{N} (1-\phi(x_{i},t_{i}))
\label{equation:sloshing:montecarlo}
\end{equation}

where $N$ {\em generated events} $(x_{i},t_{i})$ are distributed according
to joint probability density $G(x)\tilde{R}(t)$ with $\tilde{R}(t)=
R(t)/N_{0}$, $N_{0}$ being the total number of events detected during
integration time window. A list of all coordinates of the detected events
is regarded as a sample from the $G(x)$ distribution, while a similar list
of all times as the one from $\tilde{R}(t)$. Therefore, sample from
$G(x)\tilde{R}(t)$ distribution can be generated from the data by randomly
associating an event's local coordinate $x$ with another event's time $t$
among the pool of detected events. The so created coordinate-time pair is
called a generated event. The accuracy of Monte Carlo integration
increases with the number $N$ of generated events. In the time swapping
method, the function $\phi(x,t)$ defining the source region does not have
to be discretized as it had to be in the direct integration method.

Here, acceptance $G(x)$ and event rate $R(t)$ must be solutions of the
equations (\ref{equation:background:equations}) to account for on-source
events. In practice, Monte Carlo integration is performed by substituting
each real event's arrival time by a new time from the list of registered
times of collected events in a finite time window. This is why the method
is referred to as {\em time swapping method}. The swapping is repeated
$\beta$ times per each real event, $\beta$ typically being around 10. The
event rate $R(t)$ is considered to be constant on the very short time
scale and therefore is saved as a histogram. Generated event times are
drawn from it. The sample from $G(x)$ is generated using events from the
outside of the excluded region and should contain $N(x) = G(x) \int
R(t)dt$ events with given local coordinates $x$. However, the number of
events available is $N_{out}(x) = G(x)  \int \psi(x,t)R(t)dt$. Therefore,
instead of swapping each event $\beta$ times, missing events are created
by choosing actual number of swaps from a Poisson distribution with
parameter $(1+\alpha '(x))\beta$ where

\[
 1+\alpha '(x) = \frac{G(x) \int R(t)dt}{G(x) \int \psi(x,t)R(t)dt} =  
                 \frac{G(x) \int R(t)dt}{N_{out}(x)}
\]

The significance calculation has to reflect the fact that the time
swapping method is a Monte Carlo integration and thus introduces
additional fluctuations in the estimate of $N_{b}$. The integration error
reduces as the number of generated events increases or equivalently as
$\beta$ increases and the fluctuations in $N_{b}$ approach that of the
direct integration method. Use of statistic $U'$ (equation
\ref{equation:signif:notlima})  provides a transparent way of including
these additional fluctuations. It can be shown \citep{roma_thesis} that
the statistic $U'$ within framework of time swapping must be substituted
by

\begin{equation}
  U'(x)= \frac{N_{s} - N_{b}}{\sqrt{N_{s} +
         \sum_{x} \alpha(x) N_{b}(x) +  N_{b} / \beta}}, 
\label{equation:sloshing:significance_corrected}
\end{equation}
\[
  \alpha(x) = \frac{N_{b}(x)}{N_{out}(x)}
\]

The fact that the source region defined by $\phi(x,t)$ does not
have to be discretized is the advantage of the time swapping
method. Otherwise, it is based on the same assumptions as the direct
integration method: stability of the detector and isotropy of the
background (equation \ref{equation:sloshing:stability_assumption}).

\subsection{Known anisotropies.}

It was assumed in the above discussion that no anisotropy on the sky is
present. This, together with the stability assumption had lead to the
equation (\ref{equation:sloshing:stability_assumption}). In fact, if there
are known sources on the sky, then the number of registered events is
given by:

\begin{equation}
 d N(x,t) = (1 + S(x,t)) \cdot G(x) \cdot R(t) \; dx dt
\label{equation:sloshing:anisotropy_assumption}
\end{equation}

where $S(x,t)$ describes the strength of the sources as function of local 
coordinates and time.

For example, the Crab nebula is known to emit gamma rays in the TeV energy
region \citep{whipple_crab,milagro_crab}. Because the anisotropy function
$S(x,t)$ is not known, the region around the Crab has to be excluded from
the background estimation even if the nebula is not the subject of
investigation. 

Another, more dramatic example is given by two known cosmic-ray sinks on
the sky: the Sun and the Moon \citep{sun_moon}. Not only do they present a
source of anisotropy, they also traverse the sky, blocking on their way
potential candidates and perturbing on-source count $N_{s}$ as well as
$N_{b}$. This can be handled by vetoing certain size regions around the
objects, that is treating them as part of the excluded region during
integration (equations \ref{equation:sloshing:background} and
\ref{equation:background:equations}) and disregarding events if they fall
within the veto region when counting on-source events $N_{s}$. In other
words, if $\nu(x,t)$ is the function describing the veto region where it
is equal to zero and equal to one everywhere else, then excluded region
$\psi(x,t)$ and source region $\phi(x,t)$ have to be redefined as: 

\[
  \left\{
  \begin{array}{lcl}
   \psi(x,t) & := & \psi(x,t) \cdot \nu(x,t) \\
   \phi(x,t) & := & 1 - (1 - \phi(x,t)) \cdot \nu(x,t)
  \end{array}
  \right|
\]

In general, existing small scale anisotropies can be excluded or vetoed as
described, known large scale ones have to be incorporated into the
stability assumption. These will become a part of the null hypothesis
being tested.

Incorporation of the improved stability assumption
(\ref{equation:sloshing:anisotropy_assumption}) into the framework of
direct integration method is straightforward: 

\[
 N_{b} = \int \int (1-\phi(x,t)) (1 + S(x,t)) G(x) R(t) \; dxdt
\]

The anisotropy function $S(x,t)$ must be discretized on the same grid as
$G(x)$ and $R(t)$ are. In order to incorporate the improved stability
assumption into the framework of time swapping method, the generated
events $(x_{i},t_{i})$ must represent a sample from $(1 + S(x,t)) G(x)
R(t)$ which can be achieved with the help of the rejection method.

\subsection{Detector stability assumption re-examined.}

Despite the fact that no reconfigurations to the detector on the short
time scale are made, the acceptance of the array $G(x)$ depends on
transmission properties of the atmosphere which may vary during the
integration time window (equation \ref{equation:sloshing:background}). In
this case the stability assumption
(\ref{equation:sloshing:stability_assumption}) is violated and must be
replaced by assumption (\ref{equation:sloshing:anisotropy_assumption}),
where $S(x,t)$ describes the atmospheric variations. Thus, atmosphere must
be considered as an integral part of the detector and we refer to the
phenomenon in general as detector instability. If the variations are
known, they can be incorporated into background estimation as described
above. The remainder of this section presents one method of determining
such variations and shows how they are incorporated into S(x,t).

The test of the stability assumption would be a comparison of two
acceptances $G_{i}(x)$ and $G_{j}(x)$, $i \ne j$ measured at different
times $t_{i}$ and $t_{j}$. On physical grounds, a detector usually
possesses a certain degree of azimuthal symmetry, so does the atmosphere,
therefore acceptance is considered as function of zenith and azimuth
angles $G(z,A)$. The histograms $G_{i}(z,A)$ and $G_{j}(z,A)$ can be
collected from the data for a certain duration of time (for example 30
minutes) around $t_{i}$ and $t_{j}$. For the purpose of background
estimation the time scale during which the stability assumption
(\ref{equation:sloshing:stability_assumption}) holds must be ascertained. 
Therefore, the test is aimed at studying the difference between the
distributions as function of time separation $\Delta t = t_{i} - t_{j}$.

It has to be recognized that presence of sources or large scale
anisotropies on the sky and instability of the detector mimic each other,
therefore, zenith and azimuth angle distributions alone are compared
instead of two dimensional $G(x)$'s. The test can be implemented as a
series of $\chi^{2}$ tests of $G_{i}(x)$ and $G_{j}(x)$ (yielding
$\chi^{2}(t_{i},t_{j})$) and then obtaining the combined
$\chi^{2}_{total}(\Delta t)$ for time separation $\Delta t = t_{i} -
t_{j}$:

\[
   \chi^{2}_{total}(\Delta t) = \sum_{\Delta t = t_{i} - t_{j}}
                                \chi^{2}(t_{i},t_{j})
\]

The test statistic $\chi^{2}_{total}(\Delta t)$ so obtained follows a
$\chi^{2}$ distribution with $m_{total} = \sum_{\Delta t = t_{i} - t_{j}}
m(t_{i},t_{j})$ degrees of freedom if observed differences are of random
nature only. Here $m(t_{i},t_{j})$ are the number of degrees of freedom in
the corresponding $\chi^{2}(t_{i},t_{j})$ tests. The average of
$\chi^{2}_{total}$ is equal to $m_{total}$ while its variance is equal to
$2m_{total}$. Thus, the $\chi^{2}_{total}$ per degree of freedom should
fluctuate around 1.0. Examining the dependence of $\chi^{2}_{total}$ on
time separation $\Delta t$ it is possible to test the detector stability
assumption and to ascertain the proper integration time window.  If
detector instability is recognized, care must be taken to improve the
stability assumption.

\subsubsection{Illustration of diurnal modulations using Milagro.}

\begin{figure}[tbp]
\centering
\includegraphics[width=3.0in]{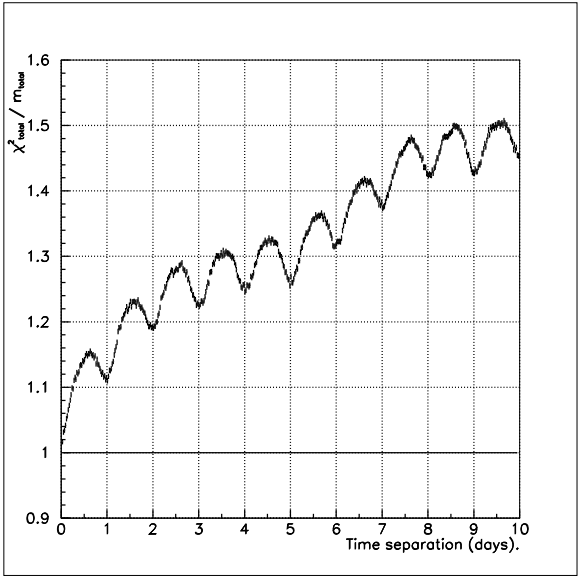}
\caption{
         Results of stability assumption test with regard to zenith
         coordinate using Milagro data. Horizontal axis is time
         separation $\Delta t$ with 30 minute bins, vertical axis is
         corresponding ${\chi^{2}_{total}(\Delta t)}/{m_{total}(\Delta
         t)}$. Solid horizontal line is the expected value of one if the
         stability assumption holds.
} \label{fig:background:stability_test}
\end{figure}

Figure \ref{fig:background:stability_test} is an example of the results of
the detector stability assumption test using Milagro data with regard to
zenith coordinate. (For description of Milagro please see
\citet{milagrito:nim}.) It is seen from the plots that the degree of
violation of the assumption grows with time separation $\Delta t$ as might
be expected, but then it drops before growing again. This can be
interpreted as presence of a periodic component which insured that two
acceptances $G_{i}(z)$ and $G_{j}(z)$ separated by 24 UT hours are
``closer'' to each other than, say, those separated by only 12. Thus,
despite the fact that no human intervention on the short time scale is
made, the acceptance of the detector changes. 

\begin{figure}[tbp]
\centering
\includegraphics[width=3.0in]{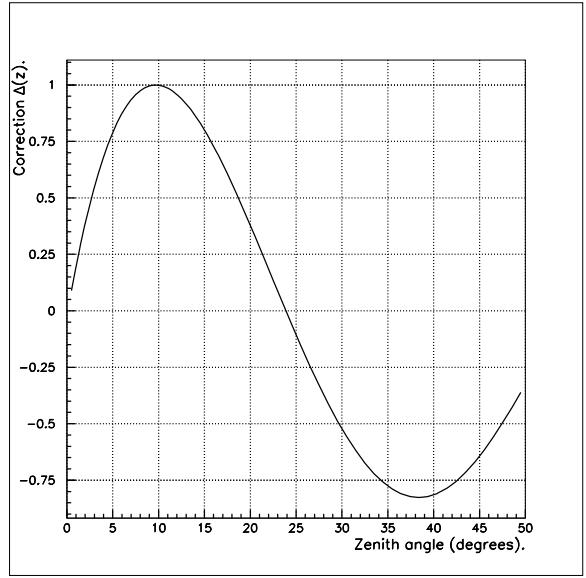}
\caption{An example of the zenith correction function $\Delta(z)$ derived
         from Milagro data.
} \label{fig:zenith_variation}
\end{figure}

\begin{figure}[tbp]
\centering
\includegraphics[width=3.0in]{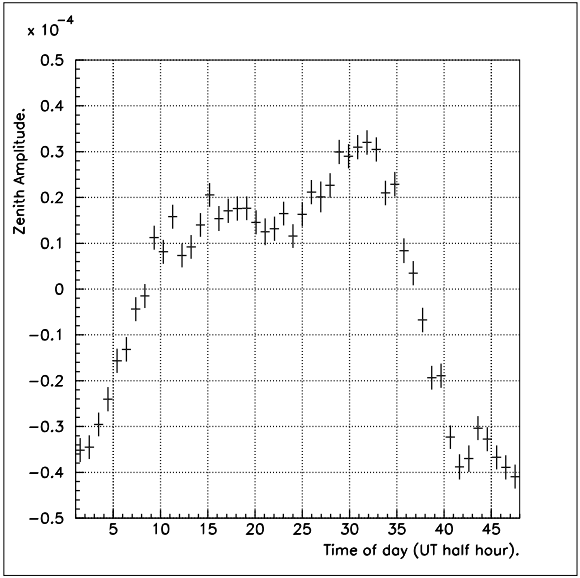}
\caption{An example of the average daily dependence of the zenith
         correction amplitude $\theta$ derived from Milagro data.
} \label{fig:backgroun:zenith_amplitude}
\end{figure}

Because the diurnal periodicity is noted, the investigation of the
modulation can be performed by comparing a particular distribution with
its daily average. It was observed that the shape of the modulation
($\Delta(z)$) of the zenith distribution is approximately constant with
amplitude varying from half hour to half hour. Therefore, the improved
stability assumption is chosen to be of the form:

\begin{equation}
  d N(x,t) = (1 + \theta(t) \Delta(z) G(x) R(t) \; dx dt
 \label{equation:sloshing:zenith_modulation}
\end{equation}

where $\theta (t)$ is the amplitude of the correction at time $t$,
$\Delta(z)$ is the polynomial zenith angle correction function
coefficients of which are obtained from the modulation shape study, $G(x)$
is the average acceptance of the detector obtained from equations
(\ref{equation:background:equations}). The example of the correction
function is shown on figure \ref{fig:zenith_variation}. The example of the
average daily amplitude dependence is shown on figure
\ref{fig:backgroun:zenith_amplitude}. The value of the amplitude is
typically within $\pm 4 \cdot 10^{-5}$ range. The plot can also be used to
justify the choice of half hour intervals for the amplitude measurement.
The assumption (\ref{equation:sloshing:zenith_modulation}) becomes part of
the null hypothesis.

\section{Conclusions.}

We have considered a typical air shower experiment conducted by means of
two observations and have discussed two commonly used tests (based on
statistics $U$ and $U'$) and conditions of their applicability. A careful
look at the situation where an astrophysical object traverses the large
field of view of a detector had led us into the subject of background
estimation. We have developed a method of background estimation which is
consistent with the use of either statistics $U$ or $U'$ and have
discussed two implementations of it: direct integration and time swapping.
The background estimation method is based on widely adopted assumptions of
short time scale stability of the detector operation and that of isotropy
of cosmic ray background. We have discussed a way to relax the short time
scale stability assumption and used Milagro data to illustrate the
situation where presence of zenith diurnal modulations can easily be
incorporated into the background estimation method. More generally, this
is also the way to incorporate known large scale anisotropies. Small scale
anisotropies do not have to be known, existing ones can be handled by
excluding or vetoing the regions around them. Any method based on the
assumption of short time scale stability of the detector operation and
that of isotropy of cosmic ray background can not be used for detection of
stationary in the field of view objects.

While the methods and ideas presented in this paper were developed for a
gamma-ray air shower array, we believe that the methods can also find
their applications outside of the field of gamma ray astronomy. The
properties of the significance test can be useful for any counting type
experiment in which number of events follows Poisson distribution, the
background estimation method can be used with any large field of view
detector, where the object of investigation traverses the field of view,
such as in solar neutrino monitors or is transient such as in SuperNovae
neutrino observatories.

\acknowledgments

We would like to thank the Milagro collaboration for permitting to use
Milagro data for illustration of the zenith diurnal modulation, and for
its help. This work is supported by the U. S. Department of Energy Office
of High Energy Physics, the National Science Foundation (Grant Numbers
PHY-9722617, -9901496, -0070927, -0070933, -0070968), the LDRD program at
Los Alamos National Laboratory, Los Alamos National Laboratory, the
University of California, the Institute of Nuclear and Particle
Astrophysics and Cosmology as well as the Institute of Geophysics and
Planetary Physics, the Research Corporation, and the California Space
Institute.

\end{document}